# Applicability of oculomics for individual risk prediction: Repeatability and robustness of retinal Fractal Dimension using DART and AutoMorph

Justin Engelmann, Diana Moukaddem, Lucas Gago, Niall Strang*, Miguel O. Bernabeu*

JE: Usher Institute, University of Edinburgh, Edinburgh, United Kingdom
School of Informatics, University of Edinburgh, Edinburgh, United Kingdom

DM: Department of Vision Sciences, Glasgow Caledonian University, Glasgow, United Kingdom

LG: Dept. de Matemátiques i Informática, Universitat de Barcelona, Barcelona, Spain

NS: Department of Vision Sciences, Glasgow Caledonian University, Glasgow, United Kingdom

MB: Usher Institute, University of Edinburgh, Edinburgh, United Kingdom
Bayes Centre, University of Edinburgh, Edinburgh, United Kingdom

*equally contributing senior authors

## Abstract

**Purpose**

To investigate whether Fractal Dimension (FD)-based oculomics could be used for individual risk prediction by evaluating repeatability and robustness.

**Methods**

We used two datasets: "Caledonia", healthy adults imaged multiple times in quick succession for research (26 subjects, 39 eyes, 377 colour fundus images), and GRAPE, glaucoma patients with baseline and follow-up visits (106 subjects, 196 eyes, 392 images). Mean follow-up time was 18.3 months in GRAPE, thus it provides a pessimistic lower-bound as vasculature could change. FD was computed with DART and AutoMorph. Image quality was assessed with QuickQual, but no images were initially excluded. Pearson, Spearman, and Intraclass Correlation (ICC) were used for population-level repeatability. For individual-level repeatability, we introduce measurement noise parameter λ which is within-eye Standard Deviation (SD) of FD measurements in units of between-eyes SD.

**Results**

In Caledonia, ICC was 0.8153 for DART and 0.5779 for AutoMorph, Pearson/Spearman correlation (first and last image) 0.7857/0.7824 for DART, and 0.3933/0.6253 for AutoMorph. In GRAPE, Pearson/Spearman correlation (first and next visit) was 0.7479/0.7474 for DART, and 0.7109/0.7208 for AutoMorph (all p<0.0001).

Median λ in Caledonia without exclusions was 3.55% for DART and 12.65% for AutoMorph, and improved to up to 1.67% and 6.64% with quality-based exclusions, respectively. Quality exclusions primarily mitigated large outliers. Worst quality in an eye correlated strongly with λ (Pearson 0.5350-0.7550, depending on dataset and method, all p<0.0001).

**Conclusions**

Repeatability was sufficient for individual-level predictions in heterogeneous populations. DART performed better on all metrics and might be able to detect small, longitudinal changes, highlighting the potential of robust methods.

## Introduction

Retinal colour fundus images are low-cost, fast-to-acquire, and non-invasive, yet provide a detailed picture of the retinal vasculature. Thus, colour fundus imaging could provide biomarkers for systemic disease [1], a field of study sometimes referred to as oculomics [2]. A particularly promising candidate biomarker is retinal Fractal Dimension (FD) which describes the complexity of the vessel structure. A less complex vasculature could indicate poorer retinal vascular health, and this in turn might correlate with vascular health elsewhere in the body. For instance, lower FD is associated with cardiovascular disease outcomes like myocardial infarction [3]–[5] and has also been studied in relation to neurovascular conditions like dementia [6], [7].

Those are exciting and promising results, but whether they can be translated into useful tools for clinical practice is still an open question. Effect sizes and increases in predictive performance over baselines using basic, easily available information like age, sex, and smoking status are typically small. Thus, for individual level predictions, retinal traits like FD would need to have very low measurement noise, yet this has to date been understudied.

Studies also often exclude a large fraction of the available images due to insufficient quality, on the order of 25-45% [3], [5], [8] in datasets like UK Biobank that were specifically collected for research. These exclusions are especially problematic for clinical applicability of oculomics. If the measurement of the retinal trait of interest, e.g. FD, fails a quarter or half of the time, then that makes it impractical. Furthermore, being older, non-White, or male increase the risk of having poor quality images [9] and thus these exclusions introduce selection bias. This means that results of existing oculomics research might not apply equally well to everyone, and if we wanted to use FD in clinical practice, the measurement would systematically fail more often for some people, e.g. those of non-White ethnicity.

Thus, we set out to investigate whether FD-based oculomics could be used for individual risk prediction by evaluating first evaluating FD's repeatability at a population- and an individual-level, without any image quality exclusions. We use two tools for computing FD: AutoMorph, which follows the established paradigm of segmentation, skeletonization, and box counting; and Deep Approximation of Retinal Traits (DART), which uses a novel paradigm of directly computing FD via a deep learning model that is trained to be more robust to image quality. We then examine how repeatability changes with the level of image exclusions due to quality and look at the relationship between measurement noise and image quality at the level of individual eyes.

## Methods

### Datasets

**Table 1**: *Overview of the datasets used, reporting statistics for all subjects we included. Interval is the time between baseline and follow-up image in GRAPE, and between first and last image in Caledonia. SD: Standard Deviation.*

|  | **Caledonia** | **GRAPE** | **Combined** |
|---|---|---|---|
| **Subjects** | 26 | 106 | 132 |
| **Eyes** | 39 | 196 | 235 |
| **Colour fundus images** | 377 | 392 | 769 |
| **Interval Mean ± SD [Min-Max]** | 2.4 ± 6.6 months [12 hours-26 months] | 18.3 ± 13.3 months [5.3-53.1] | / |
| **Female sex** | 13 (50%) | 52 (49%) | 65 (49%) |
| **Age in years Mean ± SD [Min-Max]** | 24.0 ± 3.6 [18-33] | 41.7 ± 15.0 [18-74] | 38 ± 15 [18-74] |
| **Ethnicity** | 17 White, 6 Asian, 2 Black, 1 Middle Eastern | Presumably primarily or entirely Chinese | Primarily Chinese, 17 White, 6 Asian, 2 Black, 1 Middle Eastern |
| **Ocular health** | 3 cases of amblyopia, otherwise healthy | All glaucoma (103 [97%] Open Angle; 3 [3%] Angle Closure) | 106 glaucoma, 3 amblyopia, 23 healthy |
| **Refractive status** | 12 hyperopes, 7 myopes, 7 emmetropes | Not available, but presumably mostly myopes as most subjects have Open Angle Glaucoma | Primarily myopes, at least 7 emmetropes, at least 12 hyperopes |

We included two datasets for this study: First, the "Caldonia" dataset which was collected at Glasgow Caledonian University, Glasgow, Scotland, United Kingdom. Second, the "Glaucoma Real-world Appraisal Progression Ensemble" or "GRAPE" dataset [10] which was collected at the Eye Center of the Second Affiliated Hospital of Zhejiang University, Hangzhou, Zhejiang, China. Both studies had ethical approval and adhered to the Declaration of Helsinki. Participants in both studies signed a written consent form. Table 1 shows a detailed overview of both datasets.

The Caledonia dataset was collected on a Topcon DRI OCT Triton Plus as part of a PhD project looking at choroidal thickness. Thus, the main focus was acquisition of optical coherence tomography (OCT) volume scans, but fortunately colour fundus images were acquired at the same time for most scans. Multiple scans were taken on a single day, though in some cases the data collection was repeated due to insufficient OCT quality. Thus, 5 subjects were imaged on 2 different days, 3 subjects on 3 days, and one subject on 4 days. We included every eye with at least 5 available colour fundus images. The subjects were 20 students and 6 PhD candidates at Glasgow Caledonian University.

The GRAPE dataset was collected on a Topcon TRC-NW8 (108 eyes) and a Canon CR-2 PLUS AF (88 eyes) during clinical practice. The first examination was for suspected glaucoma, with subsequent follow-up visits to monitor progression. Subjects were treated with IOP decreasing drugs after their first visit and only those with glaucoma are included in the study. We included all eyes that had a baseline and follow-up colour fundus image, taking follow-up images from the first follow-up visit that had an available image.

We analyse both datasets to examine FD in 132 subjects, imaged at two different locations with three different devices, covering a large age range, different ethnicities and both healthy and glaucomatous eyes. The Caledonia dataset provides relatively ideal conditions for repeatability, namely many images per eye, collected on the same or a handful of days in a research setting, in young adults that are generally easier to image. However, the colour fundus images were not a focus during the data collection, so the quality will likely vary at least somewhat.

The GRAPE dataset is a longitudinal dataset with only one image per eye per visit and a mean follow-up time of 18.3 months. FD a measure of retinal vascular complexity and general vascular health which could conceivably change between visits. Thus, even a perfectly repeatable method would not be expected to produce the same measurement for both visits. Furthermore, data was collected during clinical practice in a population that included over 60 year olds and thus image quality is likely more mixed. Especially as FD is calculated from the vasculature but for glaucoma the optic disc is most important, so images that were sufficient for the clinical purposes during collection might be suboptimal for calculating FD.

Based on these considerations, we expect Caledonia to provide a slightly optimistic estimate for repeatability, whereas GRAPE should provide a pessimistic, lower-bound for repeatability. Taken together, these two datasets will allow us to characterize the repeatability of FD well.

## Computation of Fractal Dimension

We used DART (short for "Deep Approximation of Retinal Traits") [11] and AutoMorph [12] to calculate FD from the colour fundus images. AutoMorph is a multi-step pipeline consisting of a deep learning model for vessel segmentation followed by skeletonization and box counting to compute FD. This is a similar approach to other tools for calculating FD like VAMPIRE [13].

DART, on the other hand, uses a single deep learning model to directly output FD from the image. DART's deep learning model was trained to replicate the output of VAMPIRE on images from UK Biobank with sufficient quality to apply VAMPIRE and achieved very high internal validity (Pearson correlation of 0.9572 on 14,907 held-out validation images). DART was trained to not just replicate VAMPIRE's output but also to be more robust to image quality. During the training progress, the model either received the original, high quality image or a poor quality version of it obtained by randomly adjusting brightness, contrast, and gamma, simulating imaging problems with anisotropic blur and gaussian noise, and adding artefacts to the images. In both cases, the model was tasked to output the FD VAMPIRE calculated from the high quality image, encouraging it to ignore variations in image quality and thus be more robust.

We chose these methods as they are both openly available on Github, allowing researchers to easily and freely access them without seeking prior permission. Furthermore, AutoMorph is a method following the traditional paradigm of segmentation, skeletonization and box counting, while DART uses a novel, yet less tried paradigm. For transparency, we want to make the reader aware that two authors of this work (JE and MB) were involved in the development of DART and thus – despite our best efforts to be neutral and objective – the reader should critically examine the present work.

## Assessment of image quality

We assessed image quality with QuickQual, a recently proposed very efficient method that leverages vector embeddings from a foundation model for natural images and obtains state-of-the-art on the EyeQ dataset. Concretely, we use the "MEga Minified Estimator" (MEME) version of QuickQual that provides a continuous "Good-Bad" quality score (probability of being bad, p(bad) for short) as opposed to EyeQ's original 3-way classification into "Good", "Useable", and "Bad" images. Images in the "Useable" class were mapped to p(bad)=0.5 during the training of QuickQual-MEME so that the model learns to put imperfect yet useable images in-between good and bad images. Thus, QuickQual-MEME's quality score is ideal for examining different levels of quality exclusions.

## Statistical analysis

### Terminology

There are many terms relating to measurement noise, e.g. "agreement", "reliability", "reproducibility", and "repeatability", that are used differently by different authors [14]. In many investigations, a key focus is to compare measurements that are made by different human annotators, something that is not applicable here as we use deterministic, fully-automatic methods. For simplicity, we use the term "repeatability" throughout while carefully explaining what data we analysed (above) and what metrics we calculate (below).

### Population-level metrics

For population-level metrics, we use Pearson and Spearman rank correlation, as well as the Intraclass Correlation Coefficient (ICC). Pearson correlation is a linear measure and sensitive to outliers. Spearman rank correlation is a robust, non-parametric measure that uses the Pearson correlation of the ranks of both variables, instead of the raw values of the variables themselves. Thus, Spearman correlation is robust to outliers and captures how similar the ranking is across both sets of measurements. Pearson and Spearman are applicable to paired measurements, e.g. two FD values of the same eye.

For more than two measurements per eye, as we have in the Caledonia dataset, we need to use the ICC instead. Though commonly referred to as *the* ICC, there are multiple versions, some of which are to examine agreement between different raters, which is not applicable here. We use the ICC as described by Bartlett & Frost [14], namely $ICC = \frac{(SD\ of\ subject's\ true\ values)^2}{(SD\ of\ subject's\ true\ values)^2 + (SD\ measurement\ error)^2}$, where SD is Standard Deviation. In words, this captures how much of the variation in the data is due to between-subject variation as a fraction of the total variation of the data. The total variation in the denominator is composed of the between-subject variation and variation due to measurement error. If measurement error was 0, the ICC would be 1. As measurement error becomes large relative to between-subject variation, the ICC decreases and approaches 0 in the limit.

The ICC is an abstract, unobservable quantity that we need to estimate based on the data. We estimate the subject's true values by taking the mean of all available images of a given eye, and then take the SD of all eyes. The SD measurement error is equivalent to the within-subject SD. Which population we choose to calculate the inter-subject variation is a key design choice that must be made taking into account the specific context, which is also stressed by Bartlett & Frost [1]. As we examine the measurement error in the Caledonia dataset, estimating the between-subject variation in the same data is the natural choice. However, as that dataset consists of healthy adults, we would expect their vascular health to be good and thus FD to have little variation. Therefore, we present the reader with two versions of the ICC. One using the inter-subject variation from Caledonia ("ICC"), and one adjusted version using the inter-subject variation from the combined Caledonia and GRAPE datasets ("Adj. ICC").

Finally, we also report Pearson and Spearman in the Caledonia dataset to enable easier comparison with the measures in GRAPE and because we expect that more readers are familiar with those measures. As we have more than two images per eye, we take two approaches to calculate Pearson and Spearman. First, we take the first and last available image per eye to make an objective, yet arbitrary choice. Second, we randomly sample one pairing per eye, calculate the correlations, and then repeat this process 20.000 times, reporting median values with an empirical 95% confidence interval. Note that bootstrapped confidence intervals for Pearson correlation can have inaccurate coverage [15]. Our sampling-based approach is similar and thus the confidence interval for Pearson might not be reliable.

*Individual-level metrics*

The metrics above summarize repeatability in a population. However, we are also interested in repeatability at an individual level. Thus, we propose the relative SD λ as a metric of individual-level measurement noise, $\lambda = \frac{SD\ of\ FD\ within\ eye}{SD\ of\ FD\ across\ eyes}$. λ expresses how large the variation of FD within an eye is compared to the variation of FD between eyes. As SD is a sum of squared mean deviations, large errors are weighted more heavily, which we think is desirable in this context. Conceptually it is similar to Pearson correlation and ICC, though for λ smaller values are better. A λ of 0 implies no measurement noise, and the larger λ gets, the more noise there is. For convenience, we express λ in %. For λ, we use the SD of FD across eyes as estimated from the combined dataset, for the reasons explained in the previous section.

*Robustness to image quality*

To examine the relationship between repeatability and image quality and to evaluate the robustness of the two methods, we first look at how λ changes in Caledonia as we exclude a larger share of images due to image quality. We consider exclusion percentages from 0% to 50%, which was chosen because it covers and spans slightly beyond typical values in the oculomics literature. Next, we relate λ and the worst image quality in a given eye. We take the worst rather than the mean quality as a single outlier could lead to a high λ. Recall that SD is based on squared differences from the mean and thus a single large deviation influences the SD more than many small deviations. We compute the Pearson correlation between λ and worst image quality, and further plot them against each other to examine the relationship between the two. This could give some insight into whether there is a critical level of quality where repeatability decreases quickly. Finally, QuickQual-MEME's quality score is the probability of an image being bad. However, probabilities are constrained quantities which can be an issue for Pearson correlation. Thus, we also evaluate the Pearson correlation between λ and the raw logit value logit(p(bad)), i.e. the raw output of QuickQual-MEME prior to applying the logistic linkage function. Note that the logistic linkage function is a monotonic and thus the Spearman correlation is the same in both cases.

## Results

### Fractal Dimension and population-level repeatability

**Table 2:** *Population-level metrics of repeatability for both methods and datasets without quality exclusions. Higher is better for all measures. 95% confidence intervals in brackets. \*\*\* denotes p<<0.0001. Note that the coverage of the Pearson correlation confidence interval for the 20.000 random pairs might have inaccurate coverage. ICC: Intraclass Correlation Coefficient, Adj. ICC: Adjusted ICC.*

|  | Caledonia | | | | | | GRAPE | |
|---|---|---|---|---|---|---|---|---|
|  | First and last image of each eye | | 20.000 random pairs per eye | | All images | | First and next visit | |
|  | Pearson | Spearman | Pearson | Spearman | ICC | Adj. ICC | Pearson | Spearman |
| **DART** | 0.7857 (0.6252-0.8825)*** | 0.7824 (0.6199-0.8805)*** | 0.6845 (0.1876-0.9483) | 0.7561 (0.5836-0.8893) | 0.8153 | 0.9907 | 0.7479 (0.6789-0.8038)*** | 0.7474 (0.6783-0.8034)*** |
| **AutoMorph** | 0.3933 (0.0888-0.6306)*** | 0.6253 (0.3859-0.7858)*** | 0.3235 (-0.0683-0.8676) | 0.6097 (0.3472-0.8043) | 0.5779 | 0.9494 | 0.7109 (0.6340-0.7740)*** | 0.7208 (0.6459-0.7819)*** |

The SD of DART FD at the eye-level was 0.00733 in Caledonia, 0.03653 in GRAPE, and 0.03557 in the combined dataset. For AutoMorph, the SDs were 0.02421, 0.08926, and 0.08841, respectively. Table 2 shows different population-level metrics for both methods and datasets.

### Individual-level repeatability

**Figure 1:** *Boxplots for individual-level measurement noise λ for both datasets and methods. The y-axis has the same scale for all boxplots in a given subplot. Subplot a) is scaled to the data range, subplot b) is scaled such the boxes themselves fit. The horizontal dashed line indicates λ =25% as a visual aid.*

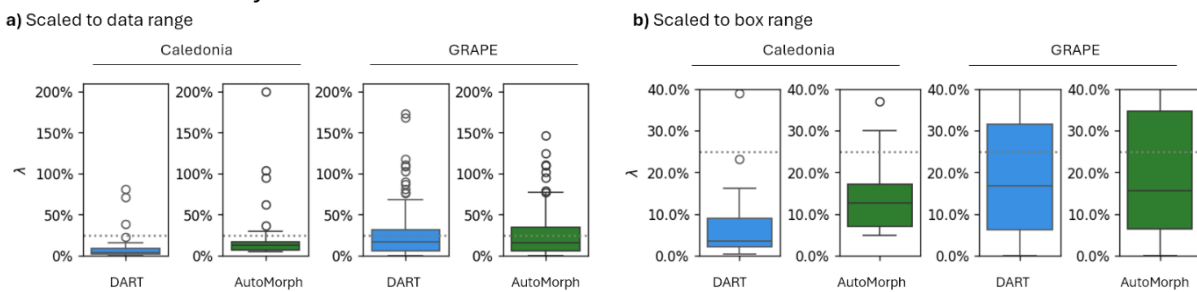

Fig. 1 shows the distributions of individual-level measurement noise λ for both datasets and methods. λ is generally higher for the GRAPE dataset, which is what we expected based on the dataset characteristics. In the Caledonia dataset the 3$^{rd}$ quantile, or 75$^{th}$ percentile, is less than 10% for DART and less than 20% for AutoMorph. For the GRAPE dataset, the 3$^{rd}$ quantiles are less than 35% for both methods.

In Caledonia, the worst values of λ were 81.02% and 199.96% for DART and AutoMorph, respectively. The image with the worst value was the same for both methods. The best values were 0.40% and 5.07%. Fig. 2 shows the best and worst examples. For the worst example, the high SD within the eye is driven by two very badly illuminated images. Removing those would change λ to 2.71% for DART and to 8.19% for AutoMorph. For the best example for DART, DART gave virtually the same FD for all images while AutoMorph had a bit more variation. On the other hand, for the best example for AutoMorph, AutoMorph has slightly less but similar variation to DART.

**Figure 2:** *Eyes with the worst and best λ for both methods. The pale boxplots in the background show the distribution of mean FD per eye across both datasets for visual reference. The coloured points with red rim show individual FD measurements for the given eye. For the worst example, we show the images for the two outliers at the bottom. All other images are randomly sampled. Below the boxplots, we indicate λ for the eye. Note that the scale of the boxplot for AutoMorph is different for the worst example as it was rescaled to fit the two outliers.*

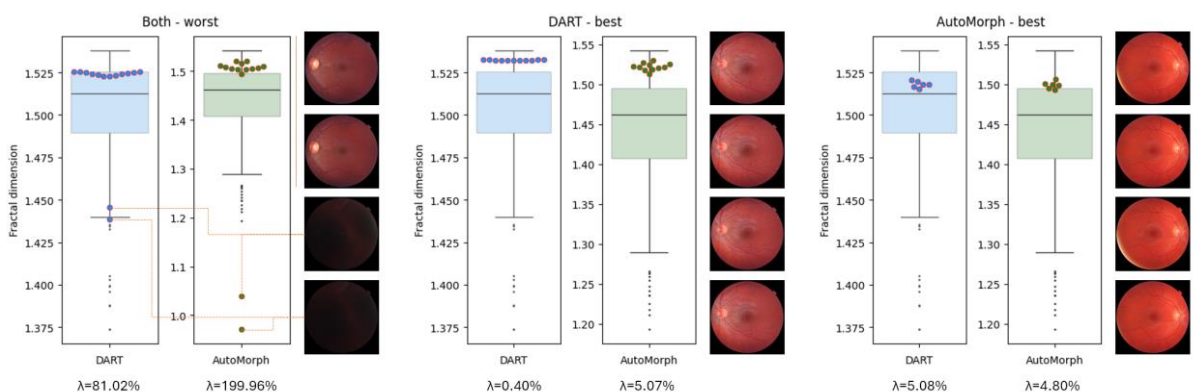

Fig. 2 also conveys a sense of how λ relates to where a subject might be placed in the distribution of FD across subjects. A λ of ~5% as in the rightmost subplot is not ideal, but ultimately places the individual in a similar location. For example, for AutoMorph all values are around the 3rd quartile and for DART between the median and 3rd quartile. However, it would preclude us from detecting differences that are around 5% of the SD of FD or smaller, e.g. in longitudinal images.

## Robustness to image quality

**Figure 3:** *Individual-level measurement noise λ versus fraction of images excluded due to quality in the Caledonia dataset for DART (a) and AutoMorph (b). Lower is better. The top, middle, and bottom numbers indicate the highest, median, and lowest values, respectively. For DART, the lowest value was a constant 0.40% and omitted for space reasons. The dashed blue line joins the highest values. Both plots have the same scaling to allow for visual comparison. When no images are excluded, there are some outliers that are denoted on the plot.*

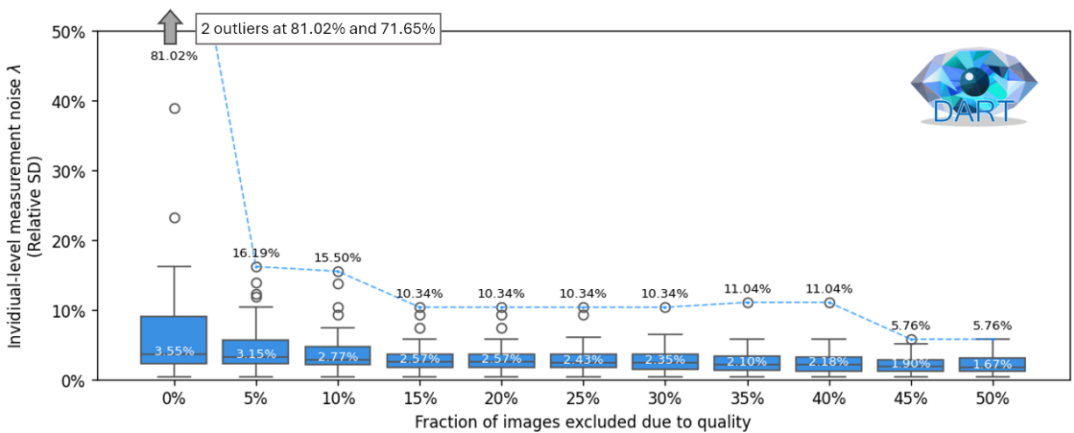

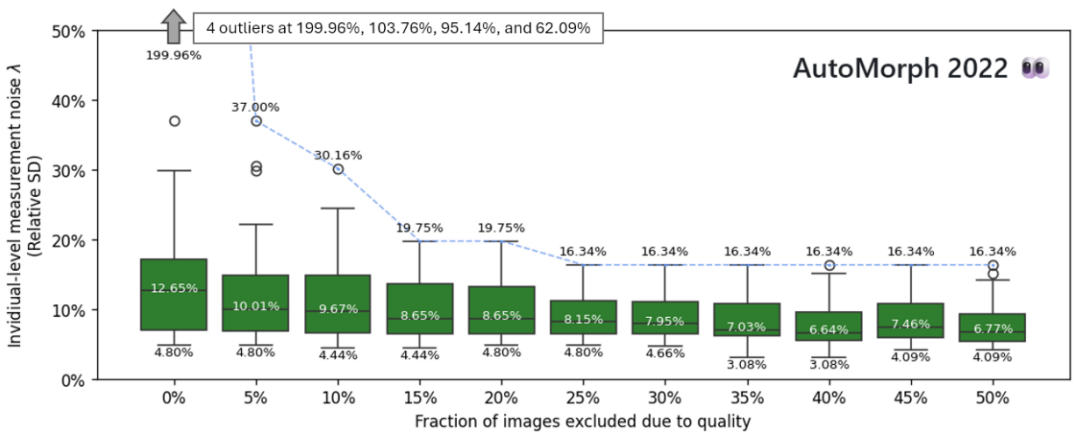

Fig. 3 shows how the distribution of λ changes as images are excluded due to quality. When no images are excluded, the highest λ for DART and AutoMorph are 81.02% and 199.96%, respectively. This decreases to 16.19% and 37.00% when the worst 5% of images are excluded. The median λ for DART is 3.55% without any exclusions, which gradually decreases to 1.67% as more images are excluded. For AutoMorph, the median is 12.65% without exclusions which decreases to 6.77% with increasing levels of exclusions.

Interestingly, the minimum λ for AutoMorph was 4.80% without exclusions, and still 3.08% with 35% of the images being excluded. This contrasts with DART, which had a constant minimum λ of 0.40% even without exclusions. Thus, AutoMorph's best case λ was 7.5-12 times higher than that of DART. AutoMorph's median was 3.5 times higher without exclusion and 3 times higher at best, namely when 40% of the images were excluded. Overall, exclusion of poor quality images primarily removes very large outliers, while median and best case repeatability only change slightly.

**Figure 4:** *Individual-level measurement noise λ versus worst quality per eye for both methods in both datasets. Each point represents one eye. All plots have the same scaling to allow for visual comparison. The Pearson and Spearman correlation between λ and worst quality is reported in the top left corner of each plot. The dashed horizontal line indicates λ=25% as a visual aid.*

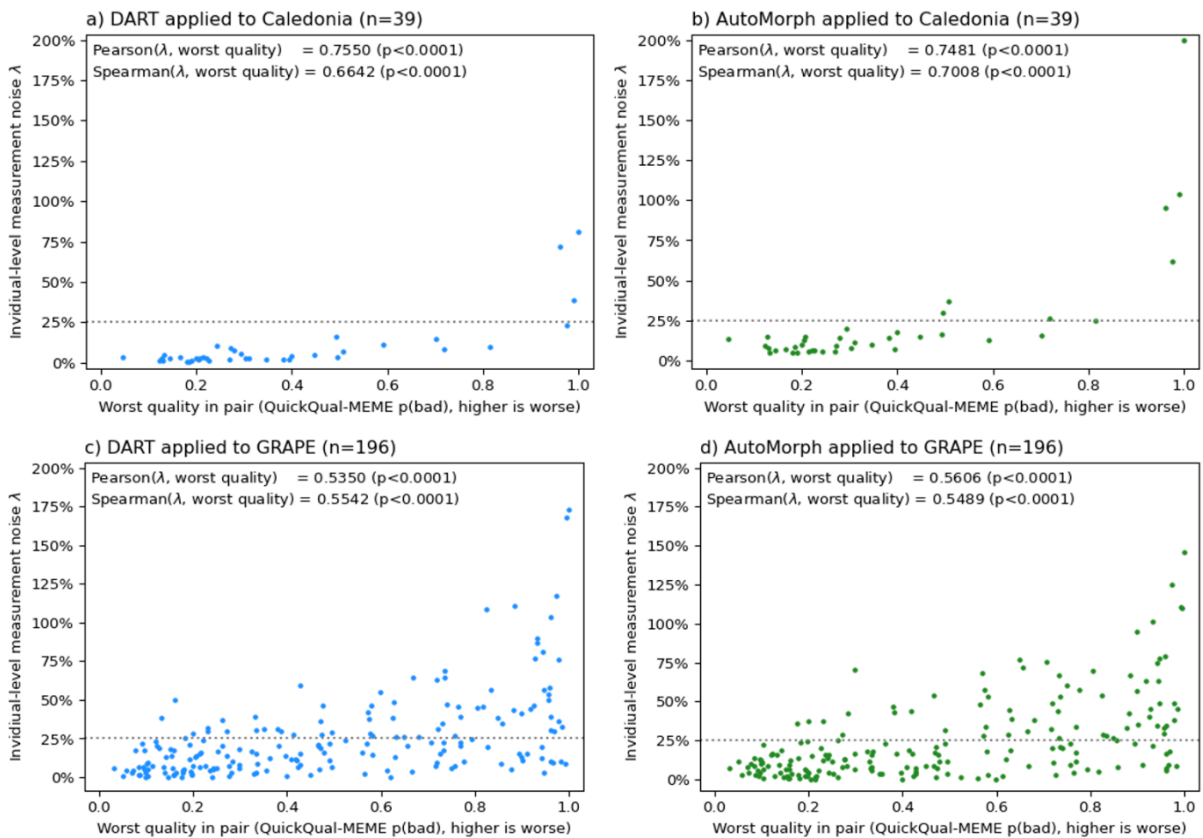

The Pearson correlation between λ and worst quality in a pair for DART was 0.7550 in Caledonia and 0.5350 in GRAPE. For AutoMorph it was 0.7481 and 0.5606, respectively. If we instead compute the Pearson correlation between λ and logit(worst quality), for DART correlations are 0.8570 and 0.5915 in the two datasets, and for AutoMorph 0.8941 and 0.6082 (all p<0.0001). Thus, the raw logits of QuickQual-MEME's quality score are a better linear predictor of λ than the probability itself.

Fig. 4 shows λ against the worst image quality in that eye. Cases of very high λ (>75%) all have very poor image quality (p(bad)>0.8), and around p(bad)=0.6 high measurement noise (>25%) appears to become more common. We can notice a visual difference between Caledonia and GRAPE. In GRAPE, there are cases of high λ even at good image quality and the correlation between λ and worst quality is lower. This is not unexpected, the long interval between images in GRAPE means that there could be due to changes retinal vasculature. However, λ is still clearly correlated with worst image quality. Thus, while there might be genuine changes in vasculature in GRAPE, cases of high λ are likely driven by poor image quality.

Discussion

Both methods showed reasonable to good repeatability at the population level, even without any images being excluded. Interestingly, Pearson and Spearman correlations were comparable between Caledonia and GRAPE, despite the fact that GRAPE should provide a pessimistic lower-bound of performance for the reasons outlined in the Methods section. This is likely due to the low between-eyes SD of FD in Caledonia as subjects were relatively young and healthy. For DART FD, SD was 5 times higher in GRAPE and for AutoMorph 3.7 times. For a constant level of absolute measurement noise, smaller between-eyes SD will yield lower correlations.

DART showed higher repeatability than AutoMorph for all metrics, especially so on the Caledonia dataset. On the GRAPE dataset, both methods were more similar. This could be due to the long follow-up time, which means that differences in FD are a combination of genuine vascular changes and measurement noise, making differences in measurement noise between the two methods appear less pronounced.

At the individual level, repeatability in terms of λ was generally good, though there were some large outliers without quality exclusions. These outliers disappear even with modest levels of image quality exclusions. Repeatability improved generally as more images were excluded, but primarily affected large outliers.

Similar to the population-level metrics, DART had smaller λs than AutoMorph, both with and without quality exclusions. Interestingly, while robustness to image quality issues was a key motivation for DART's development, DART not only had smaller outliers at low levels of exclusions, but also clear advantage in best, median and worst case λ at any level of exclusions. Thus, DART is also more repeatable in good quality images.

Based on the values of λ we observed in both datasets, both AutoMorph and DART might be applicable to individual-level risk prediction if we are targeting a population with large variation in FD, i.e. a more general population that is heterogeneous in age and systemic health. The observed values of λ are generally small enough that we would rarely confuse high-, medium-, and low-FD individuals, especially when discarding images with very bad quality, i.e. QuickQual-MEME p(bad)>0.8.

However, with median λ of 12.65% without exclusions and 6.64% with a high-level of 40% of images excluded due to quality, AutoMorph would not be able to detect small changes, e.g. in a cohort with similar age and systemic health or when looking at longitudinal changes in an individual. DART on the other hand might be able to detect such small changes with a median λ of 3.55% without exclusions and 1.67% with exclusions.

Generally, these are encouraging results for the applicability of oculomics for individual-level predictions. While population- and individual-level is a common dichotomy in the literature, a more repeatable method is necessarily less noisy and thus these results are also encouraging for population-level research. DART was more repeatable than AutoMorph even when excluding bad quality images, which highlights the value of designing robust methods for oculomics and retinal image analysis generally.

A key limitation of this work is the analysed datasets. The GRAPE dataset is longitudinal and thus only provides a pessimistic lower-bound of repeatability. On the other hand, the Caledonia dataset only contained healthy, relatively young adults and thus had low heterogeneity in FD. Additionally, there are endless alternative ways of analysing the data at hand and further metrics that readers might be interested in.

Future work should examine repeatability of FD in additional, diverse cohort. An ideal dataset for this would have a cohort spanning a wide age range, with heterogeneous systemic health, and be longitudinal with multiple images per visit, so measurement noise can be compared to longitudinal changes in the same individuals.

## Funding/Acknowledgements
The authors thank all participants in the studies used in this paper. We especially thank Prof. Kai Jin and Prof. Juan Ye as well as their colleagues for making the GRAPE dataset openly available to the research community.

JE was supported by the United Kingdom Research and Innovation (grant EP/S02431X/1), UKRI Centre for Doctoral Training in Biomedical AI at the University of Edinburgh, School of Informatics. For the purpose of open access, the author has applied a creative commons attribution (CC BY) licence to any author accepted manuscript version arising.

M.O.B. gratefully acknowledges funding from: Fondation Leducq Transatlantic Network of Excellence (17 CVD 03); EPSRC grant no. EP/X025705/1; British Heart Foundation and The Alan Turing Institute Cardiovascular Data Science Award (C-10180357); Diabetes UK (20/0006221); Fight for Sight (5137/5138); the SCONe projects funded by Chief Scientist Office, Edinburgh & Lothians Health Foundation, Sight Scotland, the Royal College of Surgeons of Edinburgh, the RS Macdonald Charitable Trust, and Fight For Sight.